\documentclass[aps,prb,twocolumn,groupedaddress]{revtex4}

\usepackage{graphicx}
\usepackage{subfigure}
\usepackage{amsmath}

\begin{document}


\title{Carrier localization mechanisms in InGaN/GaN quantum wells}


\author{D Watson-Parris}
\email[]{duncan.parris@postgrad.manchester.ac.uk}
\author{M J Godfrey}
\author{P Dawson}

\affiliation{School of Physics and Astronomy, University of Manchester, Manchester, M13 9PL, United Kingdom}

\collaboration{}
\author{R A Oliver}
\author{M J Galtrey}
\author{M J Kappers}
\author{C J Humphreys}
\affiliation{Department of Materials Science and Metallurgy, University of Cambridge,
Pembroke St., Cambridge, CB2 3QZ, United Kingdom}

\date{\today}

\begin{abstract}
Localization lengths of the electrons and holes in InGaN/GaN quantum wells have been calculated using numerical solutions of the effective mass Schr\"odinger equation. We have treated the distribution of indium atoms as random and found that the resultant fluctuations in alloy concentration can localize the carriers. By using a locally varying indium concentration function we have calculated the contribution to the potential energy of the carriers from band gap fluctuations, the deformation potential and the spontaneous and piezoelectric fields. We have considered the effect of well width fluctuations and found that these contribute to electron localization, but not to hole localization. We also simulate low temperature photoluminescence spectra and find good agreement with experiment.
\end{abstract}


\maketitle

Despite the relatively high dislocation density in InGaN/GaN quantum well (QW) structures grown on c-plane sapphire, the room temperature photoluminescence (PL) efficiency of such QWs can be very high~\cite{Nakamura:1998p3255}. This is generally attributed to localization of the carriers which reduces the effect of non-radiative recombination at the dislocations~\cite{Chichibu:1997p2272,Teo:1998p11633,Davidson:2000p248}. The precise nature of this localization is still however a matter of debate. Three possible causes of carrier localization that have been widely cited are well width fluctuations~\cite{Grandjean:2001p3716,Dhar:2002p11632,graham:103508}, random alloy fluctuations~\cite{Nguyen:2004p7935} including In-N-In chains~\cite{bellaiche:1842}, and indium clustering~\cite{Gerthsen:2000p5690,Cheng:2004p5789}. Smeeton et al. have shown however that gross indium clustering observed in TEM images of InGaN QWs can be caused by electron beam damage~\cite{Smeeton:2003p5419}.

In this work we present the results of theoretical calculations which demonstrate the importance of random fluctuations in alloy composition, and of well width fluctuations. Previous theoretical work on carrier localization has considered the properties of bulk InGaN with or without embedded InGaN quantum dots (QDs)~\cite{Wang:2001p5806,Kent:2001p5805}. Using an atomistic empirical pseudo-potential method they found that in bulk zinc-blende InGaN hole wave-functions are localised by alloy fluctuations alone, and that the electron wave-functions required a dot, or cluster like confinement to be localised. The properties of these bulk zinc-blende materials are expected to differ from those in the wurtzite QW structures we consider because of the effect of the large strain, and piezoelectric field across the QW in the case of wurtzite materials. 

Our method is to calculate the potential energy landscape of the QW using a random distribution of indium atoms. Once we have calculated this potential energy we solve the effective mass Schr\"odinger equation using the finite difference approximation to find the energy eigenvalues and wave-functions of the carriers. Performing this calculation for different distributions of indium atoms and averaging the results reveals the mean localization lengths for electrons and holes.

Experimental results showing the random nature of the indium distribution and the occurrence of well width fluctuations are described in Section~\ref{micro}. Our calculations of the carrier potential energy and the subsequent solution of the Schr\"odinger equation are described in Section~\ref{calc}, and the results of these calculations are discussed in Section~\ref{results}.

\section{Microstructural data}\label{micro}
In order to model the localization of carriers in InGaN QWs, a realistic description of their nanoscale structure is required.  Here, we base our model on the experimental data of InGaN QWs gained by Galtrey et al.~\cite{galtrey:061903,Galtrey:2008p8153} using atom probe tomography (APT).  APT has two major advantages over conventional transmission electron microscopy (TEM) studies of InGaN QW microstructures:  firstly, the three-dimensional nature of the APT data set is more appropriate to the development of a three-dimensional model than the two-dimensional projections usually recorded in TEM, and secondly the high energy electron beam used in TEM has been reported~\cite{Smeeton:2003p5419} to rapidly damage InGaN QWs, resulting in potentially unreliable structural data.  We will outline the key results of APT studies of InGaN quantum wells grown at a single temperature, and highlight the specific features of the APT dataset that we have used as an input to our model.

\begin{figure}[]%
\includegraphics*[width=\linewidth]{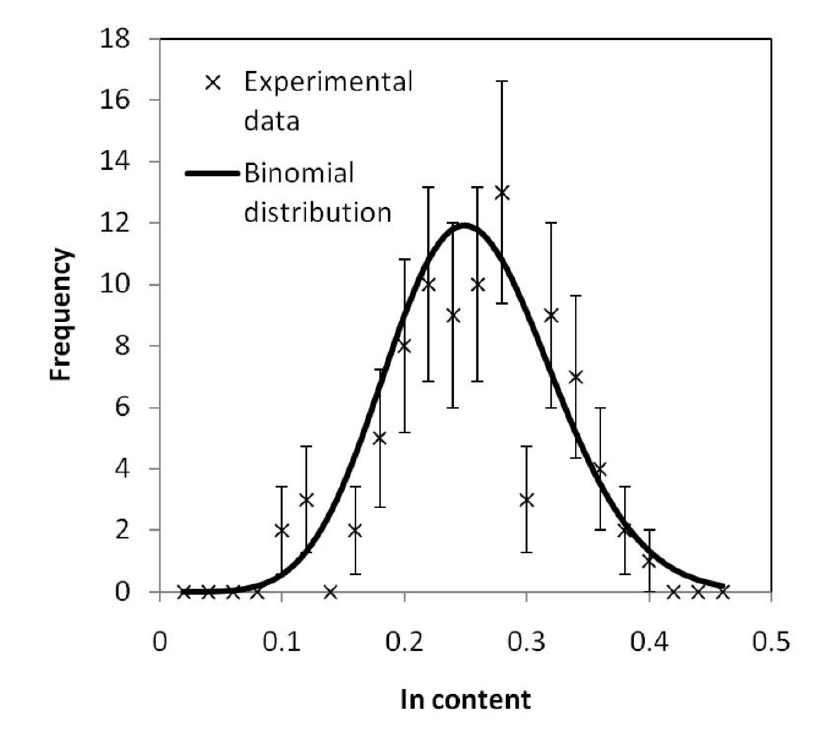}
\caption{%
Statistical analysis of the distribution of indium contents in InGaN QWs indicates no deviation from that expected in a random alloy. For the composition frequency distributions shown here, APT data from green-emitting QW material was divided into bins of 100 atoms, and the indium content was calculated for each bin. The solid line shows the binomial distribution that would be expected in the case of a random alloy. Similar data are also available in Ref~\cite{galtrey:061903}.}
\label{fig:distribution}
\end{figure}

For quantum wells grown at a single temperature, with ${x=0.18}$ and $0.25$, the APT data revealed the In$_x$Ga$_{1-x}$N within the quantum well to be a random alloy~\cite{Smeeton:2003p5419}. The experimentally-determined distribution of indium atoms within the quantum well was compared to the expected binomial distribution for a random alloy.  Figure \ref{fig:distribution} shows this comparison for a typical ${x=0.25}$ quantum well sample.  A $\chi^2$ test was used to assess whether there was any statistically significant deviation from the random distribution, but no evidence was found for any indium clustering~\cite{galtrey:061903}.  Hence, in our calculations, the InGaN within the QW has been modeled as having a random, uncorrelated indium distribution.

\begin{figure}[]%
\subfigure[]{
\includegraphics*[width=0.4\linewidth]{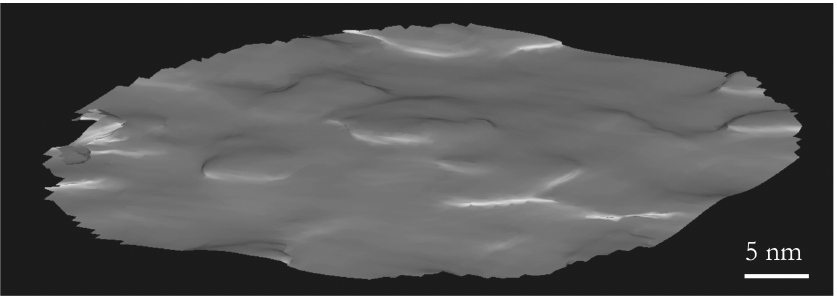}
\label{fig:3DAP}
}
\subfigure[]{
\includegraphics*[width=0.4\linewidth]{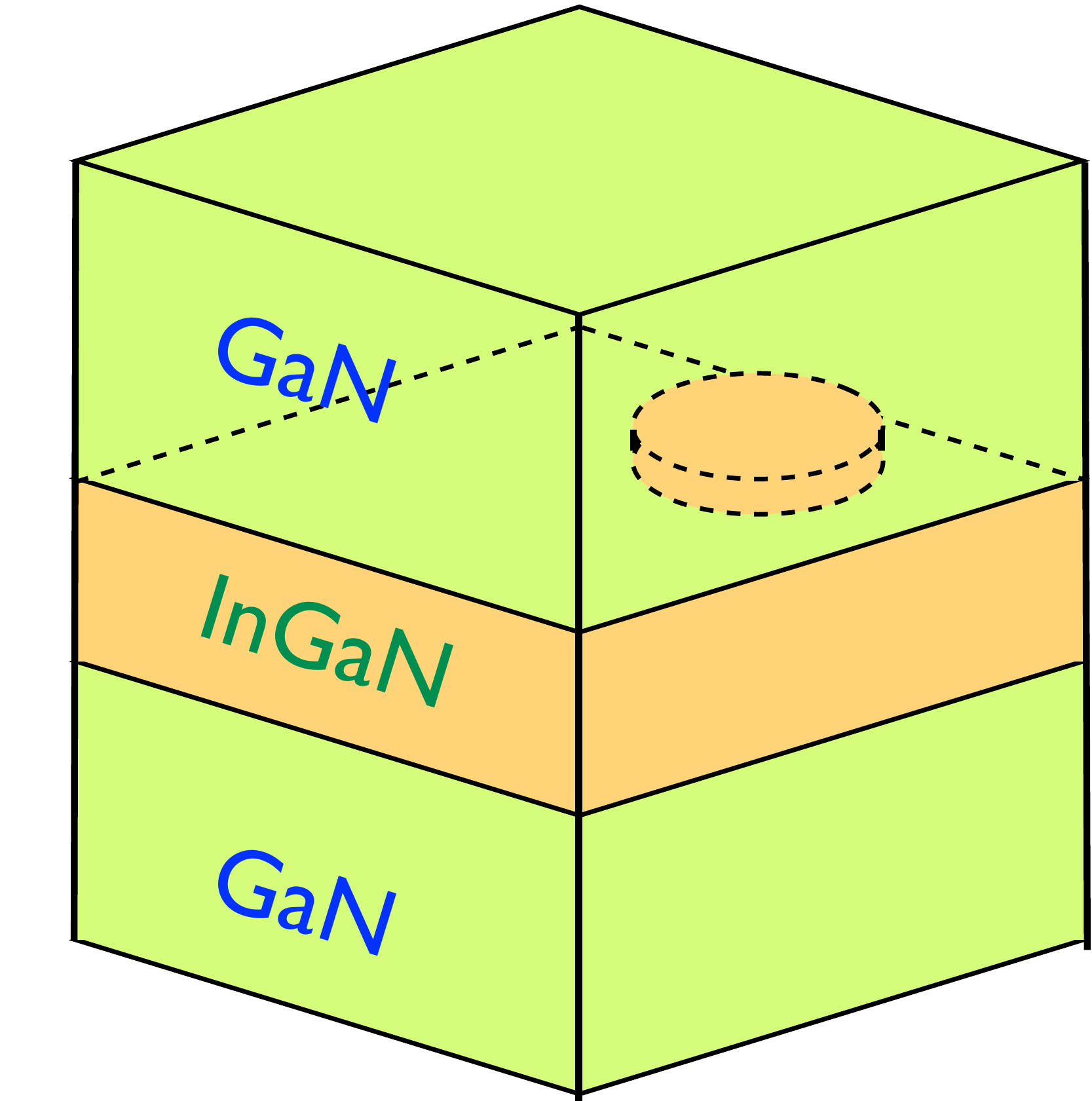}
\label{fig:schematic}
}
\caption{%
(a) An iso-concentration surface based on APT data illustrating the roughness of the upper interface of an InGaN QW.  Islands one or two monolayers high and a few nanometers across are observed~\cite{Galtrey:2008p8153}.  (b) Schematic diagram of InGaN/GaN QW with disk-shaped monolayer well width fluctuation.
}
\label{fig:WWF}
\end{figure}

Further APT studies~\cite{Galtrey:2008p8153} revealed important information regarding the two QW interfaces. We describe the interface in which the InGaN is grown on the GaN as the lower interface and the interface at which the InGaN is capped with more GaN as the upper interface. The lower interface was found to be both smooth and abrupt.  In contrast, the upper interface was both more diffuse and rougher~\cite{Galtrey:2008p8153}.  In indium iso-concentration surfaces illustrating the upper interface, such as that shown in Fig.\ref{fig:3DAP}, islands one to two monolayers high and a few nanometers across were observed on the InGaN surface, and this observation was supported by other data from TEM~\cite{graham:103508} and atomic force microscopy (AFM)~\cite{Galtrey:2008p8153}. In our model, we have included islands at the upper interface similar to those observed experimentally, since the well width fluctuations (WWFs) the islands cause have previously been predicted to significantly affect carrier localization~\cite{Grandjean:2001p3716,Dhar:2002p11632,graham:103508}. We have investigated the effect of $10$nm diameter, one mono-layer thick disk-shaped WWFs in our calculations, as schematically demonstrated in Fig. \ref{fig:schematic}. We have not included interface diffuseness in our model, but since this diffuseness is homogeneous in the plane of the QW we expect this to have a less pronounced effect on the localization than the WWFs.

\section{Theory}\label{calc}
To investigate the localization of the carriers in these structures the carrier wave functions are calculated by solution of the effective mass Schr\"odinger equation. This method requires knowledge of the potential energy for the carriers. Our calculation of the potential energy of the carriers includes the effects of the band offset between InN and GaN, the spontaneous polarization, the piezoelectric field, and the deformation potential. As this potential varies spatially in all three dimensions, this potential energy can be referred to as an ``energy landscape". We discuss the calculation of the potential energy in section~\ref{landscape} and the solution of the Schr\"odinger equation in~\ref{wavecalc}.
\subsection{Potential Energy Landscape}\label{landscape}
\begin{figure}[]
\includegraphics[width=\linewidth]{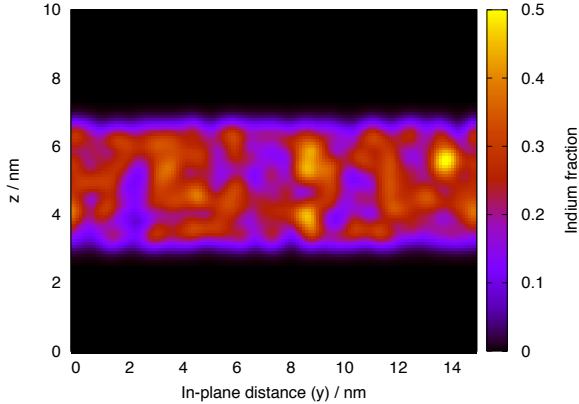}
\caption{%
A result of our calculation of the indium fraction in a QW of width 3nm with an average indium concentration of 25\%.  The z direction is normal to the x-y plane of the QW.}
\label{fig:conc}
\end{figure}

The potential energy landscape depends critically on the distribution of indium atoms. As described above we take the indium atoms to be distributed randomly in the QW. For the continuum strain mechanics we employ, the indium distribution must be described by a smoothly varying concentration function. To calculate this concentration function we first occupy cation lattice sites with indium atoms with probability equal to the nominal concentration. We use Gaussian functions with a standard deviation equal to the cation lattice spacing to smooth this positional data. An example of the resulting indium concentration function, $\chi(\mathbf{r})$, is shown in Fig.~\ref{fig:conc}. The fluctuations in indium fraction seen in Fig.~\ref{fig:conc} arise naturally from the \emph{random} distribution of indium atoms and is fundamental to the calculations which follow. It should be noted that the maxima of the indium concentration seen in Fig.~\ref{fig:conc} are \emph{not} due to individual indium atoms, but instead represent the averaged effect of several indium atoms lying close to one another; this can be confirmed by examining the distance scales on the axes of the plot.

These fluctuations in local indium fraction are responsible for the local variations in quantum confinement. To calculate this locally varying band gap we use the quadratic approximation,
\begin{equation}
\label{eq:bandgap}
E_g(\mathbf{r})=\chi(\mathbf{r})E_g^{InN} + \left(1-\chi(\mathbf{r})\right)E_g^{GaN} - b\chi\left(\mathbf{r})(1-\chi(\mathbf{r})\right),
\end{equation}
$E_g^{InN}$, and $E_g^{GaN}$, are the energy gaps in InN and GaN respectively, and $b$ is the bowing parameter. These parameters and all the relevant mechanical constants are taken, or derived, from Vurgaftman and Meyer's review article~\cite{Vurgaftman:2003p22}, and the effective mass parameters are from Rinke at al.~\cite{Rinke:2008p212}.

The strength of the spontaneous polarization which occurs in wurtzite structures also depends on the indium concentration and we describe the polarization using a linear approximation,
\begin{equation} \label{eq:Psp}
\mathbf{P}_{sp}(\mathbf{r})=\chi(\mathbf{r})\mathbf{P}^{InN}_{sp} + (1-\chi(\mathbf{r}))\mathbf{P}^{GaN}_{sp}.
\end{equation}
We can also write this as
\begin{equation}
\mathbf{P}_{sp}(\mathbf{r})=\chi(\mathbf{r})\left(\mathbf{P}^{InN}_{sp}-\mathbf{P}^{GaN}_{sp}\right) + \mathbf{P}^{GaN}_{sp}.
\end{equation}
As the polarization is purely in the z direction, we can write \(\mathbf{P}^{InN}_{sp}-\mathbf{P}^{GaN}_{sp} = \Delta P \hat{\mathbf{z}}\). This can be used in Poisson's equation,
\begin{equation}
\rho_{sp}=-\Delta P \frac{\partial \chi(\mathbf{r})}{\partial z}=-\nabla\cdot(\epsilon_0 \epsilon_r \nabla\phi_{sp}).
\end{equation}
which, after taking the Fourier transform, gives
\begin{equation}
\tilde{\phi}_{sp}(\mathbf{k})=-\frac{i k_3}{\epsilon_0 \epsilon_r k^2}\Delta P \tilde{\chi}(\mathbf{k})
\end{equation}
for the corresponding electrostatic potential. This contribution is evaluated numerically and transformed to real space using a fast Fourier transform (FFT).

Further to this the lattice constants of InN and GaN differ, so that the fluctuating indium distribution causes an inhomogeneous strain field. This strain causes a deformation potential and a large piezoelectric field. We use a Green's function method to find the strain field, and the following description closely follows that by Andreev et al.~\cite{andreev:297}. The Green's tensor $G_{kn}(\mathbf{r})$ gives the displacement at $\mathbf{r}$ in the direction $k$, due to a unit point force in direction $n$ placed at the origin. For an infinite anisotropic, homogeneous elastic medium, it is the solution to the equation~\cite{elasticity}
\begin{equation} \label{eq:Greensdef}
\lambda_{ij,kl}\frac{\partial^2 G_{kn}(\mathbf{r})}{\partial x_{j}\partial x_{l}}=-\delta(\mathbf{r})\delta_{in}.
\end{equation}
For small deformations we are then able to write an expression for the strain tensor in terms of the Green's function and concentration function, $\chi(\mathbf{r})$. We solve for the Green's function by Fourier transform and obtain an expression for the strain in Fourier space. By numerical evaluation of this function, and by performing a FFT we are able to calculate the real space strain distribution due to the inhomogeneous indium concentration function.

The deformation potential accounts for the change in energy levels due to this strain. To calculate the deformation potential the band gap energy is given in terms of  the deformation potentials, $D$ and $a$, and the strain tensor $\mathbf{U}$ by:
\begin{equation}
\label{eq:defpot}
E_g(\mathbf{U})=E_g(0)+U_{zz}\left(a_1-D_3\right)+\left(U_{xx}+U_{yy}\right)\left(a_2-D_4\right)
\end{equation}
where the $\mathbf{r}$ dependance is implicit~\cite{Wagner:2002p242}.

Once we have calculated the strain in the material finding the piezoelectric field is straightforward. The polarization caused by the piezoelectric effect can be written as $p_{i}=e_{i,kl}U_{kl}$, where $e_{i,kl}$ is the piezoelectric tensor. Poisson's equation is solved in the same manner as for the spontaneous polarization.

\subsection{Carrier Wavefunctions}\label{wavecalc}
The carrier wave functions and energy eigenvalues are obtained by solution of the effective mass Schr\"odinger equation,
\begin{equation}
\mathcal{H}\psi(\mathbf{r})=-\frac{\hbar^2}{2}\nabla\cdot\left(\frac{1}{m(\mathbf{r})}\nabla\psi(\mathbf{r})\right)+V(\mathbf{r})\psi(\mathbf{r})=E\psi(\mathbf{r}),
\label{eq:schrodinger}
\end{equation}
where we use an anisotropic, inhomogeneous effective mass. The hole in-plane (perpendicular to the c-axis) effective mass increases rapidly away from the band edge~\cite{Shields:2001p13532,Kim:1997p13975} and 1-D calculations for the c-plane QW samples detailed below revealed holes ~${\sim15-55\,}$meV away from the band edge. To avoid the use of prohibitively expensive iterative methods we have assumed far from the band edge hole effective masses as given by Chuang and Chang~\cite{Chuang:1996p218}. 

As our potential energy $V(\mathbf{r})$ has already been evaluated numerically, it is a  straightforward exercise to solve this equation using a finite difference method, with periodic boundary conditions on the wavefunction. The resulting sparse matrix eigenvalue problem is solved using the ARPACK subroutine library~\cite{ARPACK}. This allows us to find the lowest energy eigensolutions of the system.

\begin{figure}[]%
\includegraphics[width=\linewidth]{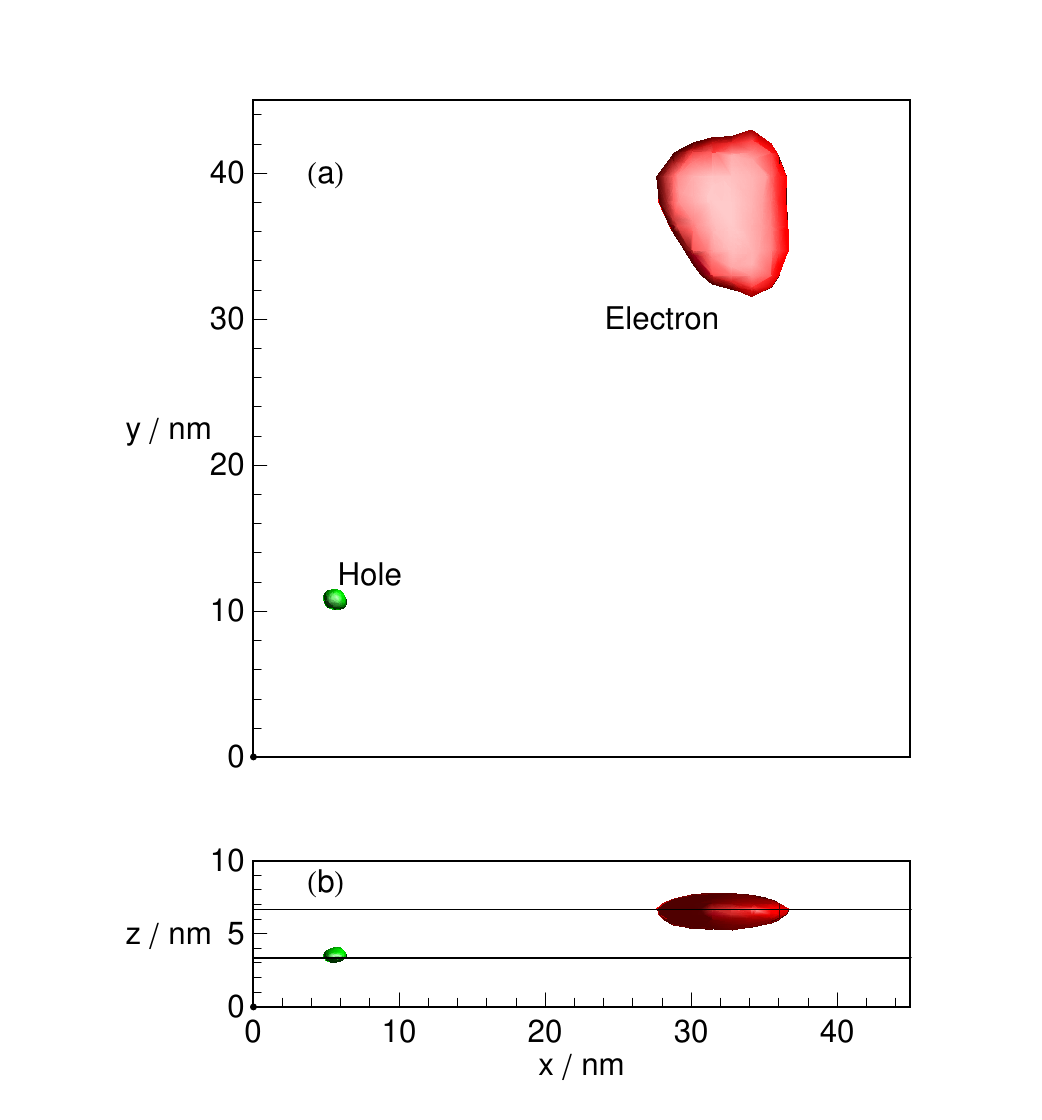}
\caption{%
The calculated ground state probability density of an electron~(red) and a hole~(green). The iso-surfaces are taken at one standard deviation of the maximum of the function, and plotted in real space in units of nanometers. Figure (a) shows the view along the z-axis, and (b) shows the view along the y-axis. The plane of the QW is normal to the z-axis, and the interfaces are at ${z=3.5}$ and $6.5$nm.}
\label{fig:wavefn}
\end{figure}
In Fig.~\ref{fig:wavefn} we show the calculated ground state probability density of an electron~(red) and a hole~(green); they were calculated for a 3nm wide 25\%-indium QW, and are clearly localised on a scale of just a few nanometers. While Fig.~\ref{fig:wavefn} shows both hole and electron wavefunctions it should be noted that the interactions between them is not included in these calculations. To quantify this localization we define our localization length as $\Delta\mathbf{r}$, using the standard quantum mechanical expression
\begin{equation}
(\Delta\mathbf{r})^2=\int\left|\left(\mathbf{r}-\left<\mathbf{r}\right>\right)\psi\right|^2\mathrm{d}^3\mathbf{r},
\label{eq:variance}
\end{equation}
where
\begin{equation}
\left<\mathbf{r}\right> = \int \mathbf{r}|\psi|^2\mathrm{d}^3\mathbf{r}.
\label{eq:exppos}
\end{equation}

\section{Results and Discussion}\label{results}
To investigate the localization properties of the carriers we performed the calculations described above for different random distributions of indium atoms. Calculation system sizes of $15\times15\times10\,$nm were used to find the 15 lowest energy hole states for 45 different distributions. The four lowest energy electron states were calculated in 65 larger $45\times45\times10\,$nm systems owing to the larger localization lengths of these states. The systems studied were 3nm thick QWs which varied in average indium composition between 5\% and 25\%. We investigate the effect of the local fluctuations in indium concentration in polar (c-plane) QWs, monolayer WWFs as described in section~\ref{micro}, and matching non-polar (a- or m-plane) samples.

\begin{figure}[]%
\subfigure[Ground-state \emph{electron} localization lengths]{
\includegraphics*[width=\linewidth]{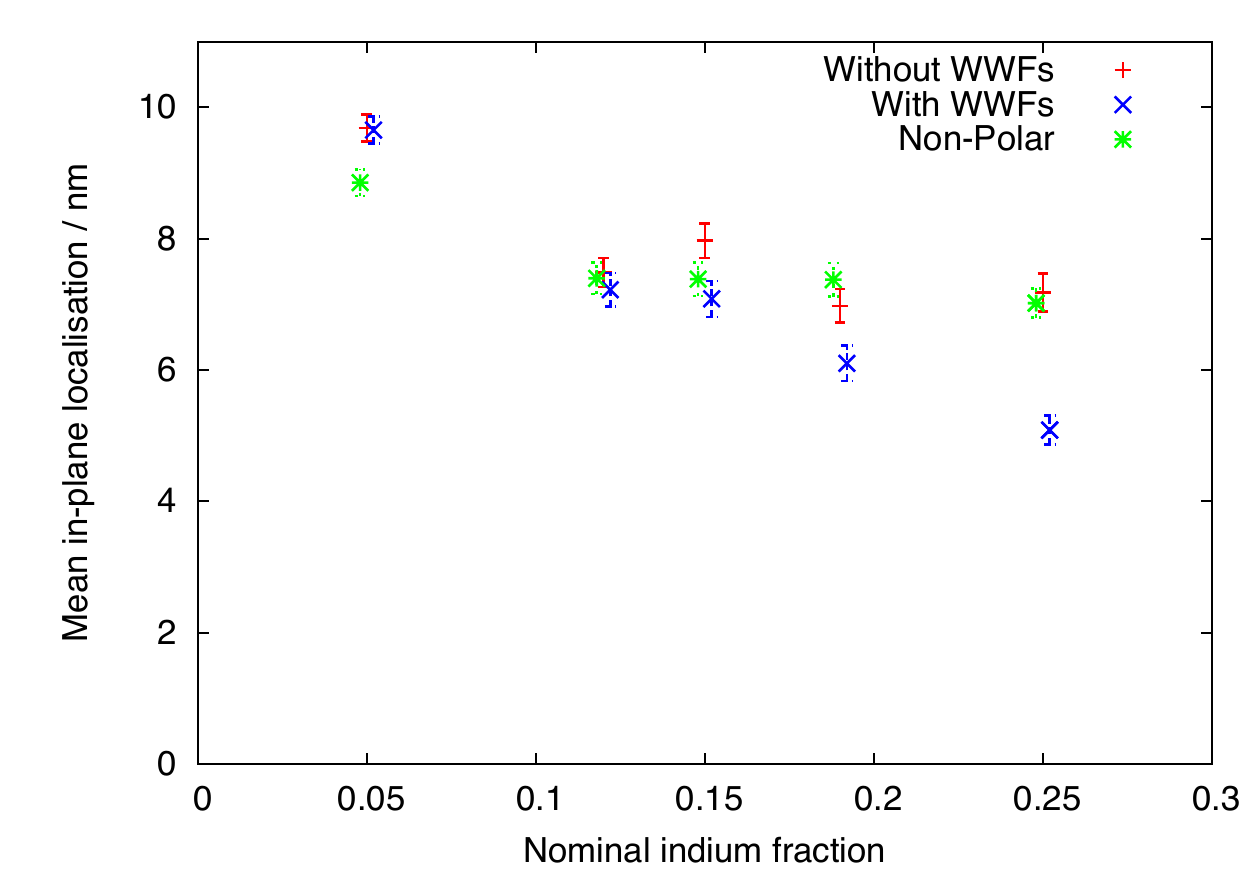}
\label{fig:hlocal}
}
\subfigure[Ground-state \emph{hole} localization lengths]{
\includegraphics*[width=\linewidth]{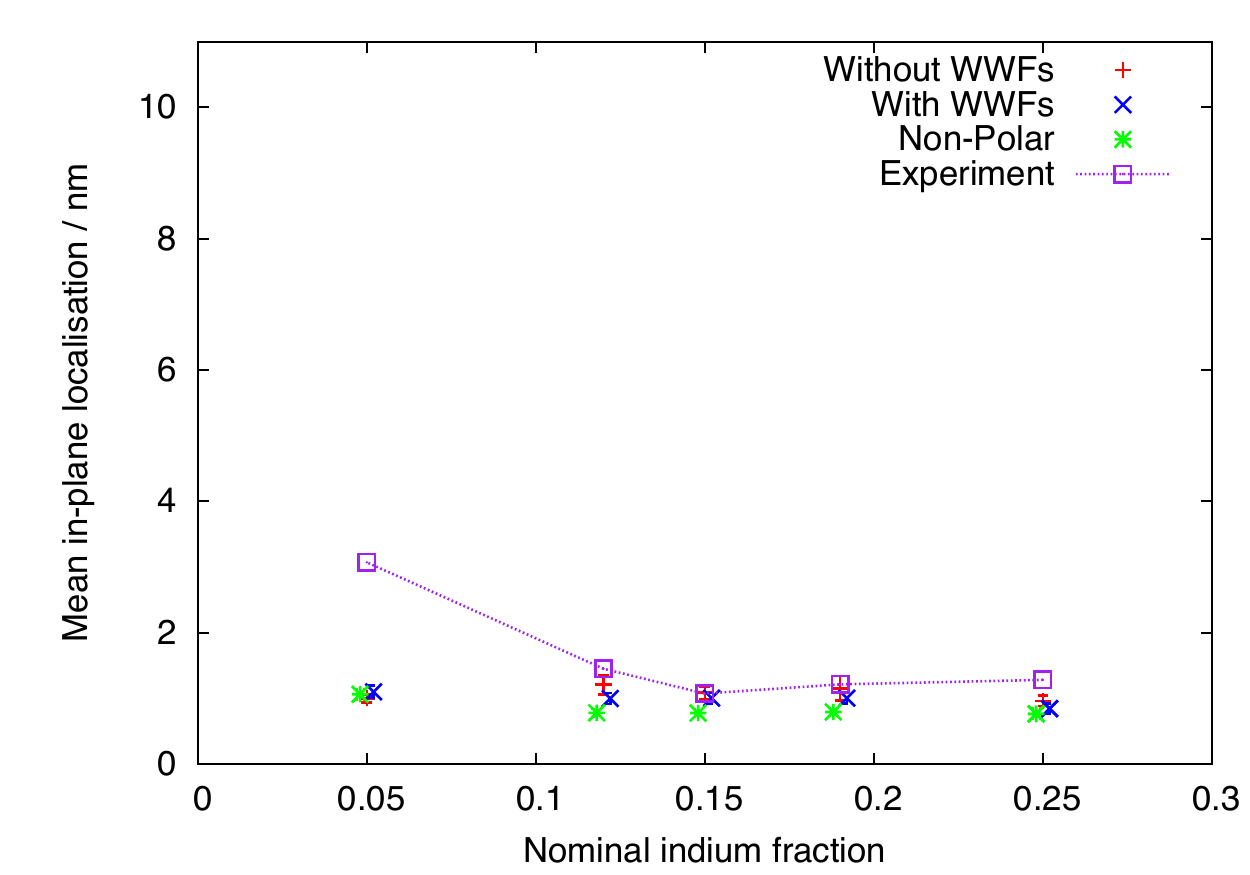}
\label{fig:elocal}
}
\caption{%
Mean ground-state localization lengths are shown for QWs of different indium fraction, with and without WWFs and in a non-polar orientation. Localization data from Graham et al.~\cite{graham:103508} are also plotted for comparison. Error bars represent the standard error.
}
\end{figure}

The mean localization lengths $\Delta\mathbf{r}$, as defined in Eq. \ref{eq:variance}, for the lowest energy hole and electron eigenstates are plotted against the nominal indium fraction in Figs~\ref{fig:hlocal} and \ref{fig:elocal} respectively. Considering the effects of indium fluctuations only, the hole wavefunctions show a constant localization length of ${\sim1.0}$nm across the range of indium concentrations. The electrons show an increasing localization length from ${\sim7}$nm for the 25\% indium sample, to ${\sim10}$nm for the 5\% indium sample. This trend of increasing localization length with decreasing nominal indium concentration is caused by the reduction in magnitude of the fluctuations (strictly the standard deviation of the concentration distribution decreases) as the nominal concentration gets smaller. The holes do not display this behavior as the larger hole effective mass means they are less sensitive to the changing depth of fluctuations. The difference in overall localization length-scale can also be attributed to the difference in the effective mass between the electrons and holes.

When we consider the added contribution of WWFs on the carrier localization lengths we see a marked difference in their effect on electrons and holes. The mean localization length for the electrons in the 25\% indium sample is reduced by a third with the introduction of this WWF, whereas for the holes the WWF has almost no effect on the localization. This difference arises from the effect of the electric field, which causes the holes to be localised on the smoother~(lower) interface, and the electrons to be localised on the rougher~(upper) interface. It should be noted that the holes are localised on a shorter length scale than the WWF.
The effect of the large electric field across the QW was investigated by comparing the polar results with calculations on QWs in a non-polar (a- or m-plane) orientation. For the polar QWs we have neglected the excitonic interaction, which is justified in wide wells due to the separation of the carriers caused by the electric field. This is not the case in non-polar QWs; nevertheless the calculations provide at least qualitative information about carrier localization in such structures. We also only consider the effect of alloy fluctuations in this orientation, not WWFs. We see that again the hole localization lengths are unchanged throughout the range of concentrations investigated. The electron localization is also unchanged, suggesting the electric field has little effect on carrier localization. At this stage we have not yet considered the effect of the electric field on electron localisation in the presence of WWFs.
We have also compared our results with the localization lengths of Graham el. al.~\cite{graham:103508} in Fig.~\ref{fig:hlocal}. They obtained Huang-Rhys factors from the strength of the LO-phonon replicas in PL spectra at low temperature and compared these with the results of calculations which assumed Gaussian wave functions for the localised carriers. This allowed them to estimate the carrier localization lengths. These show encouraging agreement with our hole localization results, which may demonstrate that the relevant localization length being probed in the experiment is that of the holes.

\begin{figure}[]
\includegraphics*[width=\linewidth]{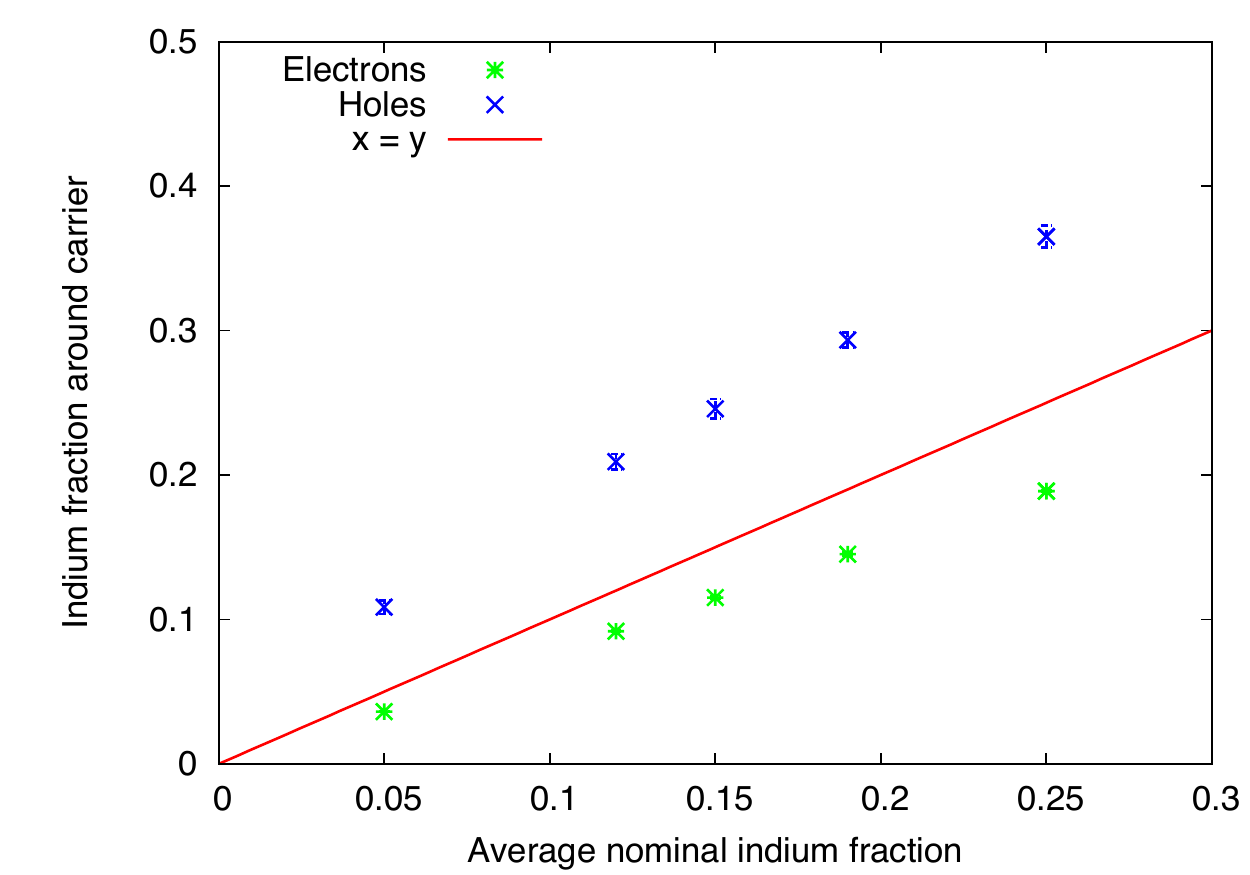}
\caption{%
The mean indium fraction within one standard deviation of the maximum of the electron and hole probability densities is plotted against the nominal indium fraction of the QW. Error bars represent the standard error in the mean.}
\label{fig:meanchi}
\end{figure}
Considering only the polar QWs with no WWFs, Fig~\ref{fig:meanchi} shows the mean indium fraction in the neighborhood of the localised carriers for each nominal concentration. We are able to calculate this by comparing the local concentration function for the sample with the calculated wave functions, and take the neighborhood of the carrier to be within one standard deviation of the peak of the wave function. Figure~\ref{fig:meanchi} clearly indicates that the holes are localised in naturally occurring regions of higher indium concentration. In the nominally 25\% sample for example, the mean indium concentration under the lowest energy hole wave functions  was 37\%.  It should once again be noted that these above average indium regions occur due to random fluctuations in composition. The mean indium concentration around the electrons in this sample was approximately 19\%. This is lower than the nominal concentration as the electron wave functions penetrate into the barrier.

We can approximate the localization energies of the carriers by fitting the density of localised states with stretched exponentials,
\begin{equation}
g_{e/h}(E) \approx n_0\exp \left[-\left(\frac{|E-E_{e/h}|}{\sigma}\right)^a\right],
\end{equation}
where ${a=1.2}$ for the holes and ${a=2}$ for the electrons. We find the localization energy ($\sigma$) of the holes in the 25\% indium c-plane sample without WWFs to be ${\sim40}$meV, and ${\sim13}$meV for the electrons. This again emphasizes the stronger localization of the holes in these structures.

All of our results point towards a picture of localization in which the holes are strongly localised by statistical fluctuations in indium concentration, and electrons are localised on a longer length-scale by a combination of these alloy fluctuations, and well width fluctuations. However including the coulombic interaction may change this picture, as the strongly localised holes may ``bind" electrons to them.

\begin{figure}[]
\includegraphics*[width=\linewidth]{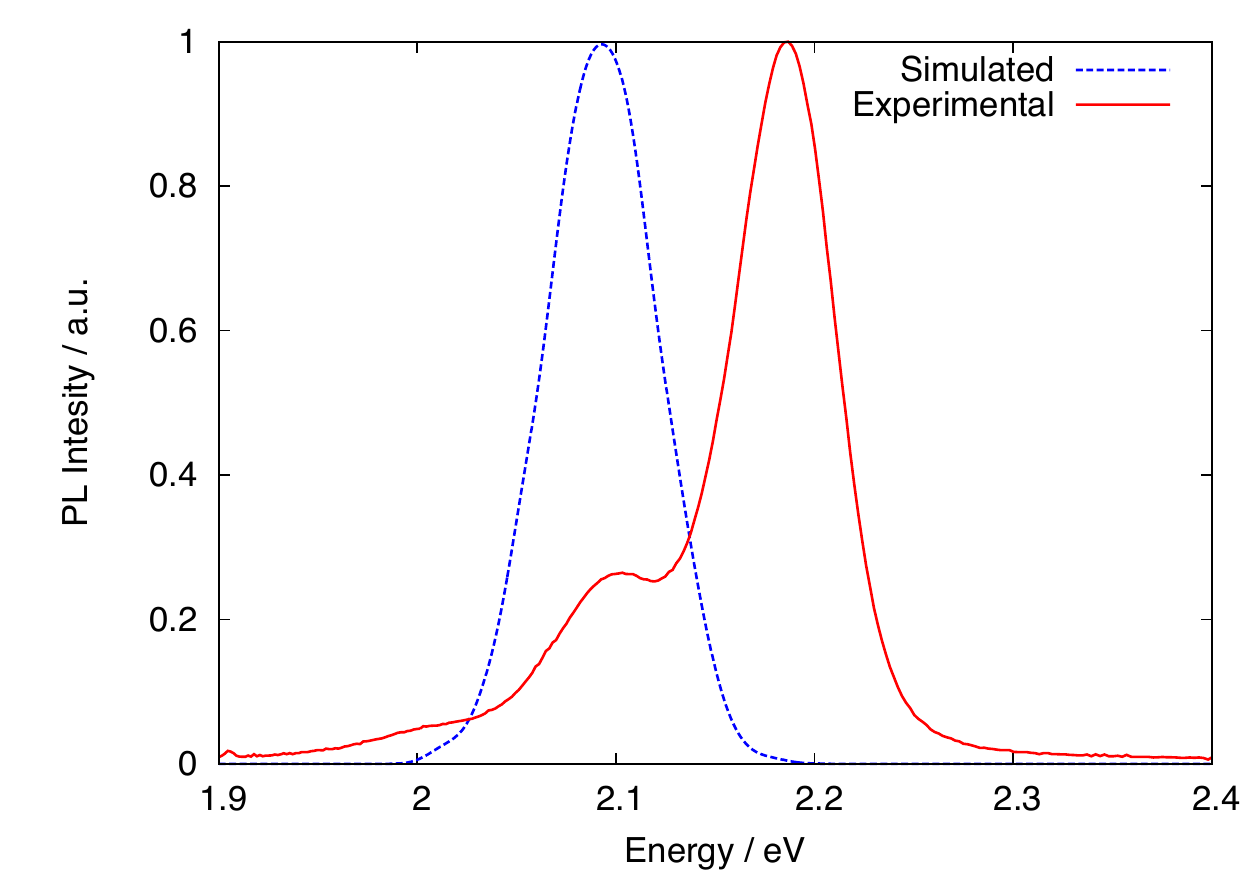}
\caption{%
A simulated PL spectrum for a 25\% indium single QW with no WWFs, compared with low temperature experimental PL spectra reported by Graham et al.~\cite{graham:103508}. The low energy feature of the experimental PL spectra are LO-phonon replicas of the main peak; the electron-phonon interaction is not included in the simulated spectrum.}
\label{fig:PL}
\end{figure}
Our calculations allow us to determine the oscillator strengths and energies of photons emitted during recombination. To ensure that we included all of the localised hole states within a given sample, we performed the calculation in a slightly smaller system, and calculated more states. We calculated 4 of the lowest energy electron states in 50 $48\times48\times10\,$nm samples, and 35 hole states in 800 $12\times12\times10\,$nm samples. We approximate the low temperature PL spectra of the samples by assuming that the localised states are all occupied by carriers, and by convolving the discrete recombinations with Gaussians of standard deviation 7meV. This width is chosen to be small enough for it not to affect the width of the final spectrum, but large enough for the spectrum to be relatively smooth. Note that due to the finite size of the system there will be long wavelength fluctuations in the indium concentration which are neglected. The effect of these would be to broaden the low energy tail slightly. Comparison of our calculated PL spectrum for a single 3.3nm thick 25\% indium QW with no WWFs with an experimental PL spectrum of a nominally identical sample is shown in Fig.~\ref{fig:PL}. The width of the peaks, often attributed to localization effects, compare very well: the simulated PL has a FWHM of 69meV, and the FWHM of the experimental PL is 63meV. In our calculations the strong hole localization is the cause of this broadening. The low energy features in the experimental PL are LO phonon replicas of the main peak and do not appear in the simulated spectrum as we have not included the electron-phonon interaction in our calculations. Including the effects of WWFs increases this broadening on the low energy side of the peak, but the effect is much smaller than that caused by the hole localization.  With allowance for the uncertainty in the experimental determination of, for example, the piezoelectric coefficients, the peak energies also agree reasonably well. 

\section{Summary}
We have been able to build a picture of the localization of the carriers in InGaN/GaN QWs by performing effective mass calculations for different random distributions of indium atoms. We have found that the holes are strongly localised in regions of above average indium content. The electrons are less strongly localised by the indium fluctuations, but become more localised by mono-layer well width fluctuations. Our work demonstrates carrier localization without the need for gross indium clusters, which goes some way to explaining the carrier localization mechanisms which are so important for the operation of InGaN optoelectronic devices.

\begin{acknowledgments}
We acknowledge the financial support by the UK Engineering and Physical Sciences Research Council, grant references EP/E035191/1 and EP/E035167/1. RAO would like to acknowledge financial support from the Royal Society.
\end{acknowledgments}


\end{document}